\documentclass[preprint,12pt, a4paper]{elsarticle}

\usepackage{amssymb}
\usepackage[breaklinks=true]{hyperref}
\usepackage{xurl}
\usepackage{enumitem}
\usepackage{booktabs}
\usepackage{longtable}
\usepackage{graphicx}
\usepackage[caption=false]{subfig}
\journal{SoftwareX}

\usepackage{amsmath}
\usepackage{standalone}
\usepackage{tikz}
\usetikzlibrary{arrows.meta, fit, backgrounds, shapes.geometric, positioning, shadows}
\usepackage{pgfplots}
\pgfplotsset{compat=1.18}
\usepgfplotslibrary{statistics}
\definecolor{brandblue}{HTML}{3366CC}
\definecolor{lightblue}{HTML}{EAF3FF}
\definecolor{outlierred}{HTML}{B32424}
\definecolor{outlierbg}{HTML}{FEE7E6}

\definecolor{buildcol}   {HTML}{D6E8FF}
\definecolor{externcol}  {HTML}{DCDCDC}
\definecolor{artifactcol}{HTML}{EAE0F5}
\definecolor{apicol}     {HTML}{B8D0F0}
\definecolor{uicol}      {HTML}{FFF3CC}
\definecolor{modelcol}   {HTML}{C4EDE7}
\definecolor{phasebg}    {HTML}{F7F7F7}
\graphicspath{{figs/}{./}}
\usepackage{cleveref}

\begin{document}

\begin{frontmatter}

\title{MediaWiki Code2Code Search: Neural Retrieval for the Semantic Discovery of Open-Source Software Entities}

\author{Francesco Tosoni}
\ead{Francesco.Tosoni@santannapisa.it}
\address{Sant'Anna School of Advanced Studies, p.zza Martiri della Libertà 33, 56127 Pisa PI, Italy}

\begin{abstract}
Code search in large-scale ecosystems is often hindered by the lexical gap between user queries and implementation details, alongside the trade-off between the low latency of traditional Information Retrieval (IR) and the precision of Deep Learning (DL). We present MediaWiki Code2Code Search, a neural retrieval system for semantic code-to-code discovery. By indexing 1.29 million structural entities (functions, types, and templates) across 2,500+ MediaWiki repositories, our system enables retrieval based on computational intent rather than surface tokens. We employ a split-build architecture, decoupling GPU-intensive offline indexing from a CPU-only serving layer; our FAISS IVF-PQ index occupies 168.6 MiB: a 96.6\% reduction compared to a flat \texttt{float32} baseline, and achieves a median query latency of 1.85 seconds on commodity hardware, satisfying the 6 GiB RAM constraint of Wikimedia Toolforge. In a blinded evaluation graded by a three-model LLM jury (inter-judge Krippendorff's $\alpha = 0.74$), our system improves ranking quality over a BM25 baseline across a 27-query benchmark, raising the jury-mean nDCG@10 (to 0.81 from 0.63), a positive and consistent gain across all judges. Gains are most pronounced in name-obfuscated and cross-language tasks where lexical methods fail. The system is available at \href{https://code2codesearch.toolforge.org}{code2codesearch.toolforge.org} under the Apache 2.0 licence and provides an open RESTful API.
\end{abstract}

\begin{keyword}
Neural Information Retrieval \sep Semantic Code Search \sep MediaWiki \sep Software Heritage \sep Vector Embeddings \sep Large Language Models
\end{keyword}

\end{frontmatter}

\section*{Current code version}

\begin{table}[!h]
\begin{tabular}{|l|p{6.5cm}|p{6.5cm}|}
\hline
\textbf{Nr.} & \textbf{Code metadata description} & \textbf{Value} \\
\hline
C1 & Current code version & v2.1.0 \\
\hline
C2 & Permanent GitHub link to code/repository & \url{https://github.com/ftosoni/mediawiki-code2code-search} \\
\hline
C3 & Legal Code License & Apache Software License 2.0 \\
\hline
C4 & Code versioning system used & git \\
\hline
C5 & Software code languages, tools, and services used & Python, JavaScript, SQLite, FAISS, Qwen3, Software Heritage \\
\hline
C6 & Compilation requirements \& dependencies & Python 3.11, Node.js (for UI), 6 GiB RAM (min) \\
\hline
C7 & Link to developer documentation/manual & \url{https://www.mediawiki.org/wiki/Code2Code_Search} \\
\hline
C8 & Support email for questions & \url{Francesco.Tosoni@santannapisa.it} \\
\hline
\end{tabular}
\caption{Code metadata. The tool is registered on Wikidata as item \href{https://www.wikidata.org/wiki/Q139251277}{Q139251277}.}
\end{table}

\section{Motivation and significance}

Wikipedia and its sister projects, attracting over 1.5 billion unique monthly visitors, are socio-technical systems underpinned by collaborative communities, open content, and open-source infrastructure \cite{rasberry_github_wikidata_2023}. Software development has been central to the Wikimedia community since its inception. The MediaWiki ecosystem (the technical backbone of most wiki projects) has evolved over two decades since 2002 \cite[\S1]{working_mw_book}. Continuous maintenance and expansion have transformed it from a monolithic application into a highly decentralised ecosystem spanning over 2,500 code repositories across various forges (e.g., Gerrit, GitHub, GitLab). This decentralisation, while mirroring the community's diversity, complicates software discovery, maintenance, and consistent data modelling. The resulting shared codebase includes toolkits for processing Wikipedia \cite{toolkit-mining-wp} and Wikidata \cite{wikidata-cacm}, bots for database analysis \cite{orcid_wd}, visualisation tools \cite{scholia_wd}, and the core infrastructure itself: MediaWiki \cite{working_mw_book} and Wikibase \cite{wd_wb_complementary}.

For maintainers, this code abundance often leads to redundant or semantically similar features. Efficient code search (retrieving relevant snippets via natural-language or code queries) is hence essential for modern development \cite{code-search-co-attentive,aroma_2019}, facilitating code completion, synthesis, traceability, and vulnerability identification.

Since 2019, the community has utilised a Codesearch tool \cite{mediawiki:codesearch} powered by Hound \cite{hound}. This tool leverages traditional information retrieval (IR) techniques, such as regular expressions and pattern matching, to index the origin/main branches of specified repositories without requiring local clones. Despite its scalability and the ``2019 Coolest Tool Award,'' the system is limited by its reliance on surface-level terminological matching. For instance, a search for ``permission validation logic'' may fail if the target function is named \texttt{checkUserLoginStatus}. This lexical gap, caused by two decades of evolving and inconsistent naming conventions, hinders volunteer onboarding and reduces efficiency for core developers.

Semantic search using deep learning (DL) models \cite{survey_llm_bigcode,aipowered-book,deep-code-search,neural-code-search,two-stage-paradigm-2023,code_search_large_scale} offers a robust alternative that balances accuracy and scalability. Consequently, the Wikimedia Foundation (WMF) and its affiliates have begun integrating AI-enabled retrieval into major projects, including Wikidata \cite{wikidatasearch}, Wikipedia \cite{wmf_ir_2026}, and Wikimedia Commons \cite{wise2023}. Notably, the Wikidata Embedding Project \cite{wikidatasearch} uses graph embeddings for about 30 million nodes, with a two-stage scheme deploying Jina AI models and a DataStax vector database. While advanced, this reliance on proprietary models and storage contrasts with Wikimedia's commitment to free, open-source software (see, e.g., the Toolforge licensing policies\footnote{\url{https://wikitech.wikimedia.org/wiki/Special:PermaLink/2239062}}). Other recent initiatives include the WMF's hybrid search models (piloted in March 2026) \cite{wmf_ir_2026} and the WISE multimodal image search engine \cite{wise2023,wise2026}, which provides semantic search for a curated subset of Wikimedia Commons images.

\paragraph{Our proposal}

Despite these advancements, the underlying codebase governing Wikimedia projects still lacks a scalable, efficient code-search implementation. To address this, we present a code-to-code retrieval system that uses code embeddings to index MediaWiki-related repositories. Our primary contributions are threefold: (i) we transform snippets into vector representations, enabling intent-based nearest-neighbour retrieval by Euclidean (L2) distance over L2-normalised embeddings; (ii) we have archived the 2,500+ MediaWiki repositories in Software Heritage (SWH) \cite[\S2.1]{swh-ecosystems-book}\cite{swh_cacm,di-cosmo-archiving,zacchiroli-archiving-repro}, ensuring a unified data model across forges and utilising citable, ISO/IEC-compliant SoftWare Hash IDs (SWHIDs) \cite{ISO-IEC-18670:2025,swhid-website}; and (iii) we provide a modern GUI built on Codex\footnote{\url{https://doc.wikimedia.org/codex/main/}} localised in 30 languages at the time of writing (14 of them are Indian) and continuously expanded by volunteer translators through the translatewiki.net platform, improving accessibility and lowering barriers for non-English-speaking contributors.

\begin{figure}[htbt]
\centering
\includegraphics[width=.9\linewidth]{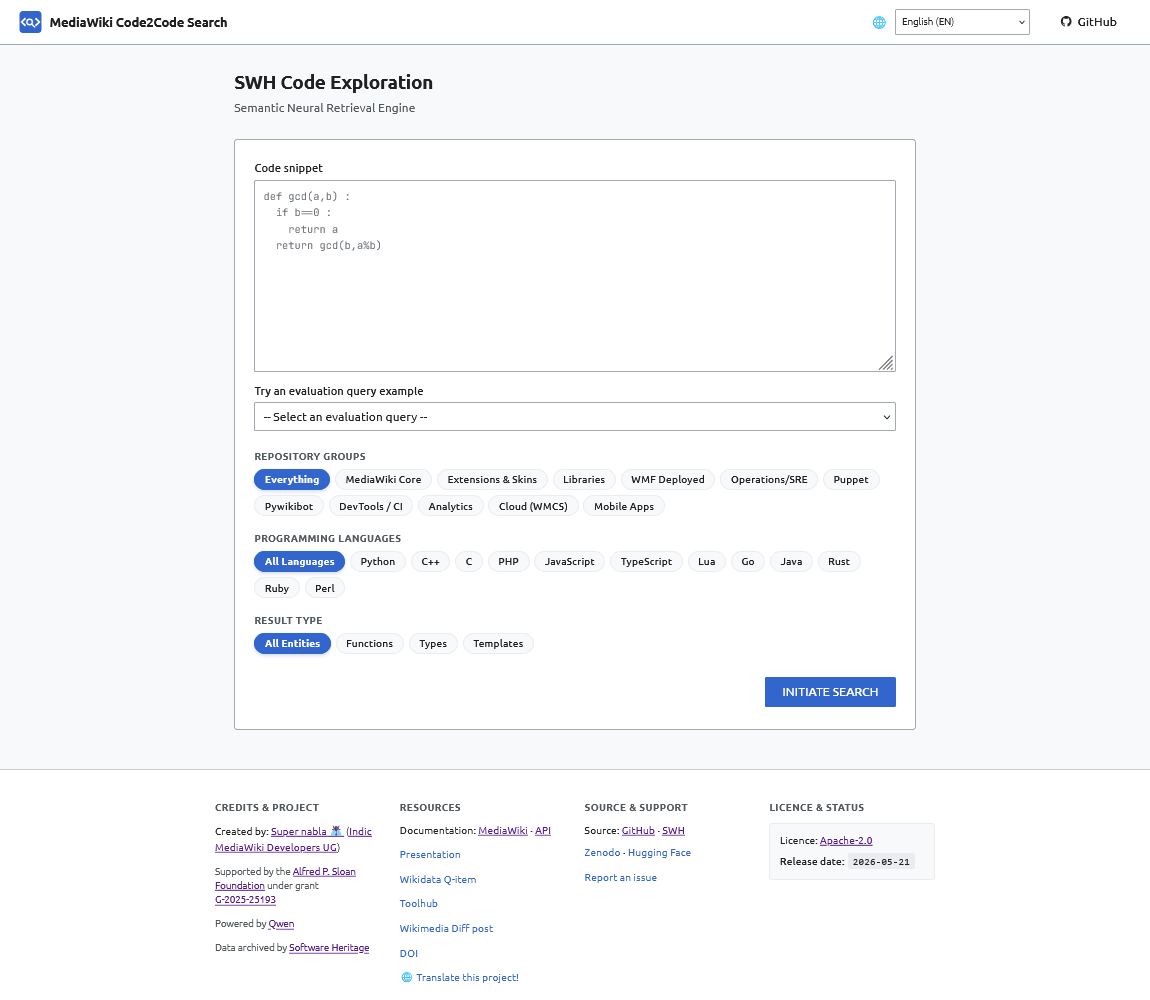}
\caption{The Codex-based frontend interface.}
\label{fig:frontend}
\end{figure}

Our architecture follows a build-heavy, serve-light approach: we use GPU-accelerated infrastructure to project code snippets into embeddings; conversely, the deployment environment remains resource-constrained and CPU-only. Unlike existing initiatives that rely on proprietary components, our entire system, including the chosen Qwen3 model, is released under the Apache Software License 2.0. The system is currently deployed on Toolforge (\url{https://code2codesearch.toolforge.org/}). It is accessible via both the Codex frontend (illustrated in \Cref{fig:frontend}) and a RESTful API with documentation served via an auto-generated Swagger UI at the \texttt{/docs} endpoint.

\section{Software description}

The implementation of the Code2Code Search pipeline is structured into discrete modules to support maintainability, CI/CD workflows, unit testing, data integrity, and resource efficiency.

A broad range of pre-trained code models has been proposed for code understanding and retrieval, and empirical comparisons show that no single model dominates across tasks \cite{comparison_source_code}. The system adopts {\tt Qwen/Qwen3-Embedding-0.6B} \cite{qwen3_emb} for its OSI-compliant licence and because a recent dedicated benchmark of deep-learning models for large-scale code-to-code recall \cite{bench_dl_code} indicates it offers a compelling trade-off between inference efficiency and code-search precision across multiple languages, making it viable for a resource-constrained environment.

\subsection{Software Architecture}

\begin{figure}[htbp]
\centering
\includegraphics[]{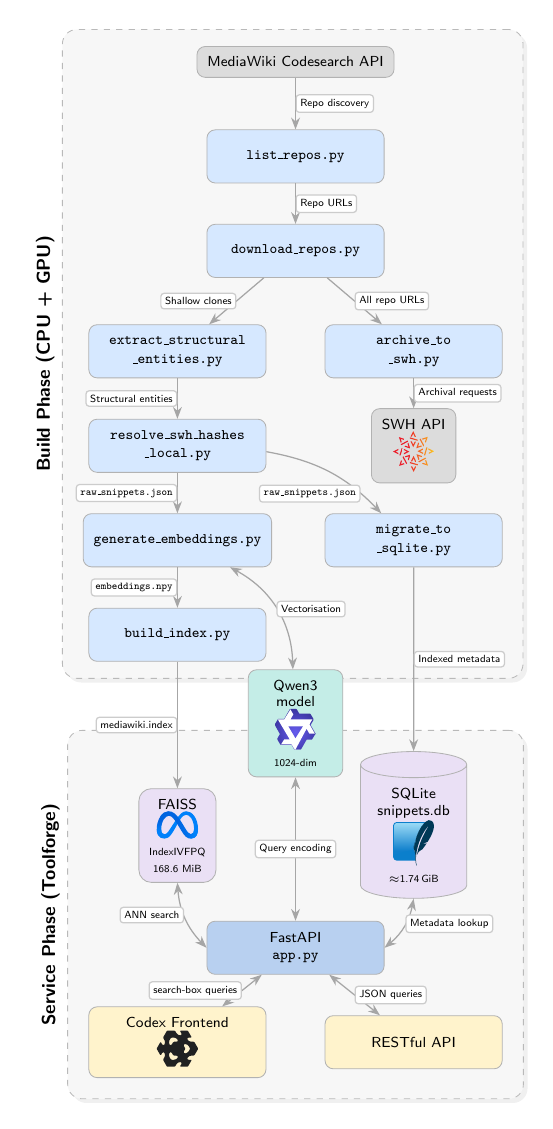}
\caption{System architecture of MediaWiki Code2Code Search.
  Node colours:
  \protect\tikz[baseline=-0.6ex]\protect\draw[fill=buildcol,   draw=black](0,0)rectangle(0.9em,0.7em);~build scripts;\enspace
  \protect\tikz[baseline=-0.6ex]\protect\draw[fill=externcol,  draw=black](0,0)rectangle(0.9em,0.7em);~external services;\enspace
  \protect\tikz[baseline=-0.6ex]\protect\draw[fill=artifactcol,draw=black](0,0)rectangle(0.9em,0.7em);~data artefacts;\enspace
  \protect\tikz[baseline=-0.6ex]\protect\draw[fill=apicol,     draw=black](0,0)rectangle(0.9em,0.7em);~API backend;\enspace
  \protect\tikz[baseline=-0.6ex]\protect\draw[fill=modelcol,   draw=black](0,0)rectangle(0.9em,0.7em);~ML model;\enspace
  \protect\tikz[baseline=-0.6ex]\protect\draw[fill=uicol,      draw=black](0,0)rectangle(0.9em,0.7em);~frontend.}
\label{fig:architecture}
\end{figure}

The system employs an asymmetric (build-heavy, serve-light) architecture, as illustrated in \Cref{fig:architecture}. Python dependencies are managed via the \texttt{requirements.txt} file at the project root.

The pipeline comprises the following stages:

\begin{enumerate}

\item \textbf{Repository Discovery:} We adopt the repository enumeration strategy of the legacy Codesearch tool. The script \texttt{list\_repos.py} queries the \texttt{/api/v1/repos} endpoint across predefined categories (e.g., \texttt{search}, \texttt{extensions}, \texttt{pywikibot}). We identified 2,568 repositories, primarily hosted on \texttt{gerrit.wikimedia.org} (1,372), \texttt{gitlab.wikimedia.org} (826), and \texttt{github.com} (342). Canonical \texttt{gerrit} URLs are prioritised by replacing read-only \texttt{gerrit-replica} addresses.

\item \textbf{Cloning:} Using \texttt{download\_repos.py}, we perform shallow clones (\texttt{--depth 1}) of the main branch for each repository. Consistent with existing infrastructure, Git submodules are excluded. Approximately 65 repositories are omitted due to duplication, authentication requirements, or download timeouts.

\item \textbf{Archival:} To ensure persistence, \texttt{archive\_to\_swh.py} submits all repositories to Software Heritage (SWH) in a single bulk request. When elevated privileges are unavailable, we provide \texttt{archive\_individual\_to\_swh.py}, which supports sequential archiving and checkpointing for resumable execution.

\item \textbf{Structural Extraction:} \texttt{extract\_structural\_entities.py} uses Tree-sitter \cite{sitter-wagner1998,sitter-comparison2023,sitter-agnostic2025} to parse source files and extract functions, types, and templates across twelve languages. We note that 261 cloned repositories yielded no valid structural entities (as they contain only documentation, translation files, or assets in non-target languages), resulting in a final index of 2,242 active repositories.

\item \textbf{Hash computation:} Subsequently, \texttt{resolve\_swh\_hashes\_local.py} generates content-based identifiers, including \texttt{sha1} and SWHIDs, enriched with metadata (origin, filepath, lines). Using \texttt{ThreadPoolExecutor} from \texttt{concurrent.futures}, we generate a \texttt{raw\_snippets.json} manifest containing all structural and repository-level metadata.

\item \textbf{Vectorisation:} \texttt{generate\_embeddings.py} executes on a GPU-enabled machine. It loads the \texttt{Qwen3-Embedding-0.6B} model via \texttt{sentence-transformers} \cite{sentence2019} to encode code snippets into normalised 1,024-dimensional vectors. This process, which takes approximately 31 minutes on a single NVIDIA H200, produces \texttt{embeddings.npy}.

\item \textbf{Index Construction:} \texttt{build\_index.py} converts \texttt{embeddings.npy} into a compressed FAISS \cite{FAISS} \texttt{IndexIVFPQ} index. Using product quantisation \cite{prodquant} with 128 sub-quantisers (8 bits/sub-vector), we significantly reduce the memory footprint to ensure compatibility with Toolforge's constraints.

\item \textbf{Metadata Migration:} \texttt{migrate\_to\_sqlite.py} transforms \texttt{raw\_snippets.json} into an SQLite database (\texttt{snippets.db}). By decoupling metadata storage from the FAISS index, this step runs independently and stays well within the 6 GiB RAM limit.

\end{enumerate}

\Cref{tab:corpus} summarises the indexed corpus. The pre-computed artefacts produced by this offline phase (\texttt{embeddings.npy}, the FAISS index, and \texttt{snippets.db}) are archived on Zenodo (\href{https://doi.org/10.5281/zenodo.20586256}{10.5281/zenodo.20586256}) and mirrored as a browsable dataset on Hugging Face (\url{https://huggingface.co/datasets/ftosoni/mediawiki-code2code-search}), so the serving layer can be reproduced without rerunning the GPU-bound stages. Following the offline build phase, the lightweight serving layer is deployed via a FastAPI backend, providing both a web interface and a RESTful API.

\begin{table}[!h]
\centering
\begin{tabular}{lrr}
\toprule
\textbf{Category} & \textbf{Count} & \textbf{Share} \\
\midrule
Functions & 1{,}050{,}748 & 81.5\% \\
Types (classes, structs, interfaces, enums) & 237{,}653 & 18.4\% \\
C++ Templates & 1{,}051 & 0.1\% \\
\midrule
\textbf{Total indexed entities} & \textbf{1{,}289{,}452} & \\
\midrule
\multicolumn{3}{l}{\textit{Top file extensions}} \\
JavaScript (.js) & 464{,}791 & 36.0\% \\
PHP (.php) & 312{,}369 & 24.2\% \\
Go (.go) & 305{,}591 & 23.7\% \\
Python (.py) & 112{,}946 & 8.8\% \\
TypeScript (.ts) & 36{,}362 & 2.8\% \\
Other & 57{,}393 & 4.5\% \\
\bottomrule
\end{tabular}
\caption{Indexed corpus statistics (snippets.db, May 2026).}
\label{tab:corpus}
\end{table}

\paragraph{Latency}

We measured end-to-end query latency by issuing 27 benchmark queries (see \Cref{sec:examples}) to a local API instance and repeating the full run 7 times. We employed an ASUSTeK VivoBook X1605VA laptop (Windows 11), equipped with a 13th-generation Intel\textregistered\ Core\texttrademark\ i9-13900H processor (14 cores / 20 threads, up to 5.4 GHz boost), 32 GB of DDR4-SDRAM, and an integrated Intel\textregistered\ Iris\textregistered\ Xe Graphics GPU (128 MB dedicated VRAM). No discrete GPU was used.

\begin{figure}[htbt]
\centering
\includegraphics[width=.95\linewidth]{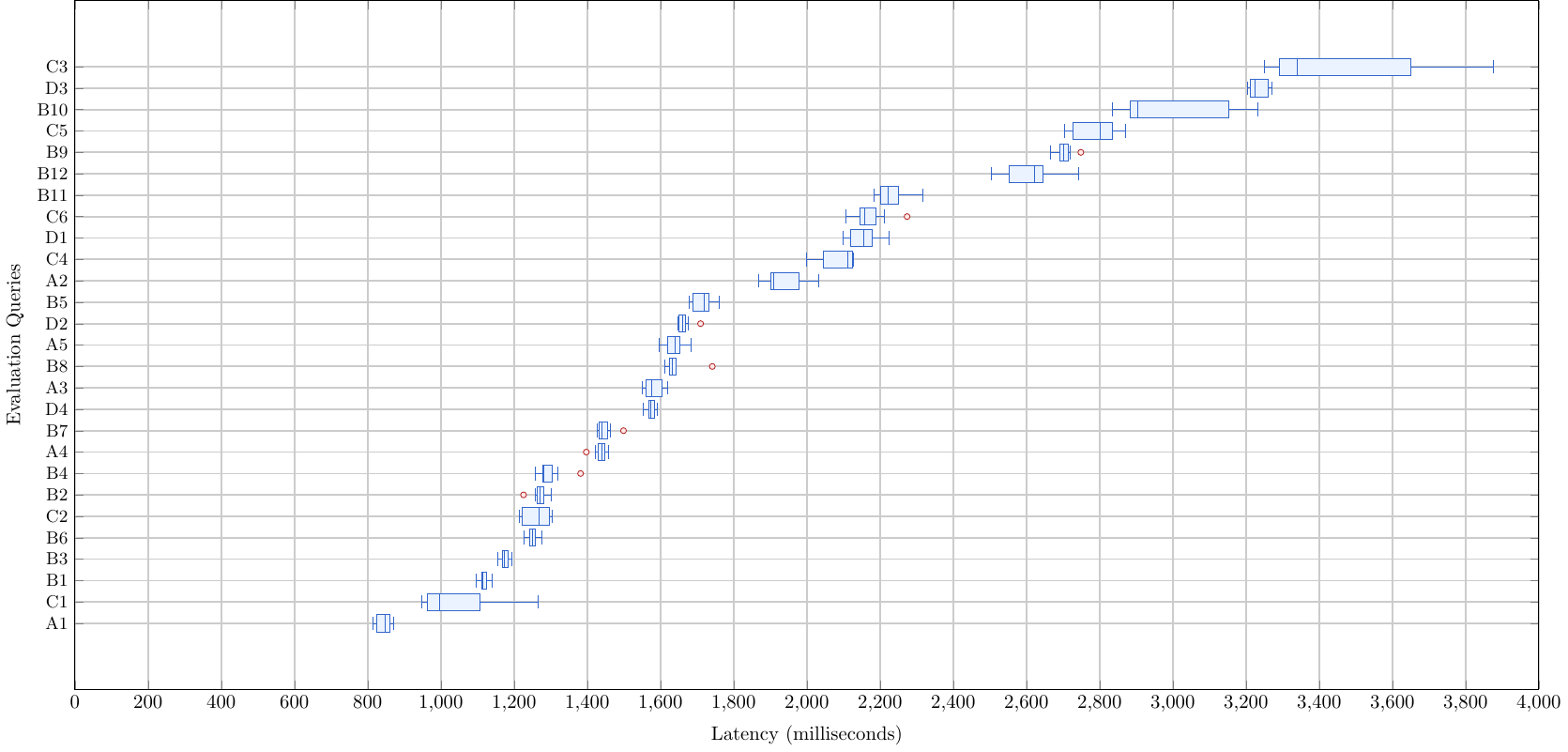}
\caption{Per-query end-to-end API latency distribution on the commodity laptop across the 27-query benchmark, measured on commodity CPU hardware (7 independent runs). Queries are sorted by median latency. Latency scales with query snippet length. On the current Toolforge deployment (not illustrated in the figure), the average latency is 31 s, with interquartiles of 19--36 s.}
\label{fig:latency_boxplot}
\end{figure}

\Cref{fig:latency_boxplot} shows the per-query latency distributions.
Averaged over the 27 queries, the per-query median latency is 1.85 s (mean 1.87 s).
Latency correlates with query snippet length, ranging from 0.9 s for the shortest query (A1, a 4-line GCD) to 3.3 s for the longest (C3, a 19-line fan-out).
On Toolforge (the live deployment, CPU-only shared virtual machines, 6 GiB RAM limit), end-to-end latency is substantially higher, averaging $\approx$31 s, ranging from 11 s for the shortest query (A1) to 111 s for the slowest (B5), because the Qwen3-Embedding-0.6B model runs on throttled shared CPU resources. GPU nodes are not currently available, but GPU deployment is the subject of an ongoing infrastructure request to the WMF to bring back inference to the order of seconds.

\subsection{Software Functionalities}

\paragraph{Syntactic Entity Extraction}

Using Tree-sitter, the system performs structural analysis across 12 languages (PHP, Python, C++, C, JavaScript, TypeScript, Lua, Go, Java, Rust, Ruby, and Perl). The extraction logic in \texttt{extract\_structural\_entities.py} employs a recursive scope-resolution algorithm to capture qualified identifiers (e.g., \texttt{Namespace\textbackslash{}Class::Method}). The frontend includes multi-select filters (by repository group, language, and entity type) to refine search results.

\paragraph{Single-Stage Neural Retrieval}

User queries are encoded at inference time using the \texttt{Qwen3-Embedding-0.6B} model, following the bi-encoder (dual-encoder) retrieval paradigm, in which queries and code are embedded independently into a shared space and compared using vector similarity \cite{sentence2019,code_search_python_translation}. Corpus embeddings are stored in a FAISS \texttt{IndexIVFPQ} index, configured with 128 sub-quantisers (8 bits per sub-vector), which trades exact search for the recall/latency profile characteristic of large-scale approximate vector similarity search \cite{hnsw,pan2024vector-db,pitfalls_vector_sigmod26}. This quantisation reduces the memory footprint by 96.6\% compared to a flat \texttt{float32} baseline (168.6 MiB versus approximately 4.9 GiB), ensuring the serving layer operates comfortably within Toolforge's 6 GiB RAM limit: an instance of the model and index-efficiency techniques increasingly required to deploy deep models under tight resource budgets \cite{gaurav-survey2023}.

\subsection{Software Verification and Technical Integrity}

System robustness and semantic accuracy are validated by a suite of 34 \texttt{pytest} unit tests. Software provenance and citability are maintained via \texttt{codemeta.json} \cite{codemeta} and \texttt{CITATION.cff} files, which are synchronised with the release SWHID and commit hash using \texttt{update\_release.py}. Further, the GitHub Actions CI pipeline (\texttt{.github/workflows/python-ci.yml}) enforces PEP 8 compliance via \texttt{flake8} and validates the parser logic on every commit; it also enables continuous deployment, as each push to the main branch triggers an automatic redeployment of the live Toolforge service: this occurs regularly, since translatewiki.net merges new and updated translations directly into the main branch.

\subsection{Portability and Reuse}

Software Heritage’s content-addressed data model \cite{swh_cacm} serves as a foundational enabler, allowing repositories from disparate forges to be mapped to a unified SWHID namespace. This architecture mitigates the challenges of multi-forge diversity and demonstrates scalability across varied software ecosystems.

The system is highly modular: only the repository discovery script, \texttt{list\_repos.py}, is ecosystem-specific. Retargeting the system for different forges, package registries, or domains (such as research institutions or SMEs) requires replacing only this single component. All downstream stages (Tree-sitter parsing, SWHID resolution, neural vectorisation, FAISS indexing, SQLite migration, and the FastAPI serving layer) are domain-agnostic. They depend solely on local source-file clones and the generated \texttt{raw\_snippets.json} manifest.

\section{Illustrative examples}
\label{sec:examples}

The frontend exposes 27 queries as ready-to-run examples in a dropdown selector (\Cref{tab:query_taxonomy}); designed with increasing difficulty, they offer a readily available sample of the codebase's functionality.

No formal ground truth exists for code search in the MediaWiki ecosystem, unlike general-purpose benchmarks such as CodeSearchNet \cite{codesearchnet} and xCodeEval \cite{xCodeEval}; nor is a direct comparison with the Hound-based Codesearch tool meaningful, given fundamentally different query interfaces (regular expressions versus code snippets).

\subsection*{Methodology}

Each of the 27 queries is a self-contained code snippet submitted verbatim to the API. For every query, we pool both systems' top-10 results (standard pooling practice \cite{voorhees1998trecpool}), deduplicate identical snippets \emph{within} and \emph{across} systems by a content hash (one shared label per snippet), and shuffle the merged pool with a per-query reproducible seed. Each candidate is graded against the relevance scale of \Cref{tab:relevance_scale}. Crucially, the judge sees only this anonymised pool: opaque identifiers carrying no system identity, rank, score, or filename. We then compute all metrics downstream in Python.

\begin{table}[!h]
\centering
\caption{Graded relevance scale applied by each blinded judge.}
\label{tab:relevance_scale}
\small
\begin{tabular}{clp{7.6cm}}
\toprule
\textbf{Score} & \textbf{Label} & \textbf{Criterion} \\
\midrule
1.0 & Relevant           & Implements the same computational task, or a correct specific solution for the query's intent. Naming, style, and \emph{programming-language} differences do not reduce relevance: a snippet solving the same task in another language scores 1.0. \\
0.5 & Partially relevant & Topically related but a partial or non-idiomatic realisation, or where the target logic appears only as a helper step within a larger function. \\
0.0 & Irrelevant         & Does not address the task, or overlaps only superficially (e.g.,\ shared identifiers while doing something unrelated). \\
\textit{null} & Indeterminable & Too truncated or context-dependent to judge; used sparingly. \\
\bottomrule
\end{tabular}
\end{table}

From these labels we compute, entirely in code, our primary rank-aware metric (nDCG@10~\cite[\S8.4]{ir_book}), lenient precision P@$k$ for $k \in \{1,3,5,10\}$ (fraction of the top-$k$ whose median label is $\geq 0.5$), a strict P@10 counting only fully-relevant results (median label $=1.0$), and MRR. Every metric is reported \emph{per judge} and as the jury mean, each with a query-level 95\% confidence interval (CI; mean $\pm 1.96$ standard errors) for both systems and for the paired Code2Code$-$BM25 difference. One caveat guides interpretation: nDCG@10 is self-normalised within each system's own retrieved top-10 (its ideal ordering is taken over exactly those items), so it captures how well each system ranks what it returned rather than coverage of the corpus; the \emph{gap between the two systems}, not the absolute score, is therefore the meaningful signal.

\paragraph{Baseline}

We compare against Okapi BM25~\cite{robertson1994okapi}\cite[\S11.4.3]{ir_book}\cite[\S3.2.1]{aipowered-book} over the same 1.29 M entities: for each query we extract its non-trivial identifiers (dropping keywords and tokens shorter than three characters) and submit them as a bag-of-words query to a BM25 index over the \texttt{code} field of \texttt{snippets.db} (the same snippet text used to build the neural embeddings).

\paragraph{Annotation protocol}

We adopt the `LLM-as-judge' paradigm, which approaches human--human agreement rates when strong models are used~\cite{zheng2023judging}. Each judge grades only the blinded, pooled, shuffled candidate set and never sees system identity, rank, or the competing ranked list. To guard against single-model idiosyncrasy, adjudication uses a \emph{jury of three judges from different model families} (Claude~Opus~4.8, Google~Gemini~3.8~Flash, and Moonshot~Kimi~K2), each queried once, with pinned model identifiers for reproducibility; the Claude judge runs via isolated sub-agents (one per query, each reading only its own blinded prompt). This iteration deliberately uses no human annotators (a cost/time trade-off); we thus report no agreement against a human gold standard and quantify reliability through \emph{inter-judge} agreement (Krippendorff's $\alpha$ of 0.74, interval). We release the blinded prompts, per-judge labels, and analysis code in \texttt{scripts/evaluation}.

\begin{table}[!h]
\centering
\caption{Query taxonomy (27 queries across four categories).}
\label{tab:query_taxonomy}
\small
\begin{tabular}{clp{7.2cm}}
\toprule
\textbf{Cat.} & \textbf{Difficulty} & \textbf{Rationale} \\
\midrule
A & Canonical algorithms       & Textbook algorithms (GCD, binary
    search, \ldots). Establishes an upper bound for both systems. \\
B & Name-obfuscated semantics  & Same computational intent as a canonical
    pattern, but expressed with non-standard identifiers. \\
C & Cross-language              & Query and relevant results are often in
    different programming languages. \\
D & Domain-specific/niche      & Patterns endemic to the Wikimedia
    ecosystem. \\
\bottomrule
\end{tabular}
\end{table}

\subsection*{Results}

\Cref{tab:eval_aggregate} reports retrieval quality per judge and the jury mean: Code2Code improves over BM25 on every metric and for every judge.

Our headline metric is the rank-aware nDCG@10, on which Code2Code raises the jury mean to 0.81 from 0.63 for BM25. A 95\% CI over the 27 queries gives $[0.73, 0.89]$ for Code2Code versus $[0.51, 0.75]$ for BM25, with the interval on the paired improvement excluding zero under all three judges. Lenient precision moves the same way (jury P@5 0.59 vs 0.43, P@10 0.51 vs 0.38), and top-rank precision follows (P@1 0.73 vs 0.56, MRR 0.80 vs 0.64), though at $n=27$ these last are high-variance, to be read as supporting evidence rather than standalone claims. Strict P@10 is low for both systems (0.21 vs 0.15), as expected from the demanding rubric.

By category (jury-mean nDCG@10), the improvement concentrates where lexical matching is weakest: Code2Code leads decisively in name-obfuscated category~B (0.83 vs 0.53) and cross-language category~C (0.81 vs 0.62), modestly in canonical category~A (0.90 vs 0.82), and trails slightly in niche category~D (0.64 vs 0.69), where domain-specific lexical retrieval remains competitive.

\begin{table}[!h]
\centering
\caption{Retrieval quality per judge and as the jury mean, over the 27-query benchmark (point estimates; higher value in each BM25/C2C pair is in bold). P@10$^{s}$ is strict P@10; nDCG is nDCG@10.}
\label{tab:eval_aggregate}
\setlength{\tabcolsep}{3.8pt}
\begin{tabular}{llrrrrrrr}
\toprule
\textbf{Judge} & \textbf{Sys.} & \textbf{P@1} & \textbf{P@3} & \textbf{P@5} & \textbf{P@10} & \textbf{P@10$^{s}$} & \textbf{nDCG} & \textbf{MRR} \\
\midrule
Claude & BM25 & 0.59 & 0.54 & 0.50 & 0.46 & 0.20 & 0.71 & 0.69 \\
Opus 4.8                & C2C  & \textbf{0.78} & \textbf{0.73} & \textbf{0.67} & \textbf{0.60} & \textbf{0.29} & \textbf{0.87} & \textbf{0.85} \\
\midrule
Gemini & BM25 & 0.48 & 0.41 & 0.36 & 0.28 & 0.11 & 0.52 & 0.55 \\
3.8 Flash                 & C2C  & \textbf{0.63} & \textbf{0.57} & \textbf{0.47} & \textbf{0.39} & \textbf{0.14} & \textbf{0.72} & \textbf{0.69} \\
\midrule
Kimi & BM25 & 0.59 & 0.49 & 0.45 & 0.41 & 0.15 & 0.66 & 0.67 \\
K2        & C2C  & \textbf{0.78} & \textbf{0.72} & \textbf{0.64} & \textbf{0.53} & \textbf{0.19} & \textbf{0.85} & \textbf{0.86} \\
\midrule
{\bf Jury} & BM25 & 0.56 & 0.48 & 0.43 & 0.38 & 0.15 & 0.63 & 0.64 \\
{\bf mean}                   & C2C  & \textbf{0.73} & \textbf{0.67} & \textbf{0.59} & \textbf{0.51} & \textbf{0.21} & \textbf{0.81} & \textbf{0.80} \\
\bottomrule
\end{tabular}
\end{table}

\subsection*{Discussion}
\paragraph{Where Code2Code excels}

Code2Code's advantage is largest where the query shares no discriminative tokens with the target: the six biggest per-query gaps (jury nDCG@10 $\Delta$ from $+0.53$ to $+1.00$) are all of this kind. In C1 (SHA-256 file hashing), the frequent token \texttt{filepath} pulls BM25 toward path utilities (\texttt{parseFile}, \texttt{validateFilePath}; 0.0 for every judge), whereas Code2Code returns the actual hashers \texttt{safeFileHash}/\texttt{getFileHash} at ranks~1--2, fully relevant to all judges. The pattern recurs in B6 and B4 (BM25 0.0) and extends to the cross-language analogues inaccessible to a token matcher, driving its category~C lead.

\paragraph{Where the two systems tie}

On about a quarter of the queries, the two systems converge, typically when the query retains discriminative library or API tokens that both methods can exploit. Seven are effective ties ($|\Delta \text{nDCG@10}| < 0.05$): A2, A4, A5, B3, B12, C2, and C4; most sit near the ceiling (nDCG@10 $\geq 0.92$), the exceptions being A4 and B12, where both are only moderately effective.

\paragraph{Where BM25 wins}

Five queries favour the baseline, though only C5 decisively: there the Lua-specific tokens \texttt{table}, \texttt{pairs}, and \texttt{tostring} let BM25 retrieve accurate recursive dumpers (nDCG@10 of 0.89), while Code2Code drifts toward unrelated Wikibase value formatters (0.42). The other four are narrow edges ($|\Delta| \leq 0.27$): B2 (debounce) and B9 (Go worker pool), where Code2Code returns behaviourally adjacent helpers rather than the exact pattern; D2 (git/SWHID hashing), coverage-limited because the distinctive \texttt{blob~\textbackslash0} framing is absent from the corpus, so both fall back to generic SHA helpers; and D4 (below). These mark the limits of dense retrieval: domain vocabulary shared by adjacent but irrelevant functions can mislead embedding similarity (the textual bias documented for dense code retrievers \cite{sacl-textual-bias}), motivating future hybrid ranking.

\paragraph{Judge disagreement and the value of a jury}

Because the judges independently grade the same blinded pool, the design surfaces genuine ambiguity rather than hiding it. D4 (MediaWiki hook dispatcher) is the clearest example: BM25 returns the exact target \texttt{HookContainer::run} at rank~1 (relevant to all), while Code2Code returns hook-execution \emph{test} helpers (\texttt{assertThatHookIsExcutable}); Claude and Kimi rate these partially relevant (0.5; nDCG@10 near 0.98), and Gemini rates these irrelevant (0.0). The jury thus records a BM25 win driven by a legitimate disagreement (whether a test harness counts as a partial answer), that a single judge would have silently resolved either way.

\paragraph{Threats to validity}

Three limitations bound our claims. \emph{(i)~Benchmark size.} At $n = 27$, the sample size, not the metrics, limits statistical power and widens the top-rank CIs; we therefore lead with nDCG@10 and treat P@1/P@3/MRR as directional support. \emph{(ii)~LLM judges.} Evaluation reliability hinges on blinding, three independent model families, and inter-judge agreement ($\alpha = 0.74$), though we make no use of human-level adjudication. \emph{(iii)~Cross-language relevance.} Counting a same-task snippet in another language as fully relevant reflects the tool's intended use but structurally advantages semantic retrieval, and part of the category-C gap follows from this declared choice. Deduplication of identical snippets removes the double-counting, though near-duplicate variants remain.

\section{Impact}

MediaWiki Code2Code Search introduces semantic retrieval capabilities to the Wikimedia open-source ecosystem. By indexing 1.29 million code snippets, the tool enables ecosystem-scale empirical research, including analysis of code reuse and duplicate functionality \cite{sourcerercc,bigclone}, vulnerability tracing (via the SWH graph \cite{swh_dag}), and alignment of best practices with actual implementations.

The system is deeply integrated into the Wikimedia technical landscape: it is registered on Toolhub, documented on MediaWiki.org, featured on the Wikimedia Diff blog, and mapped to structured metadata on Wikidata (\texttt{Q139251277}). For new developers, the tool mitigates the ``cold start'' problem; instead of navigating an unfamiliar 2,500-repository corpus via keyword search, they can identify relevant functions, classes, and templates by providing a code sketch. Experienced developers benefit from granular frontend selectors that let them filter precisely by entity type, language, and repository group.

Reflecting the WMF’s commitment to equitable access, the interface is localised through the community translation platform translatewiki.net.\footnote{\url{https://translatewiki.net/wiki/Translating:MediaWiki_Code2Code_Search}} Beginning with 17 languages (14 of them Indian), the catalogue is now maintained and extended by volunteer translators (30 languages at the time of writing, and growing), reaching from widely-spoken national languages to regional and minority ones (e.g.,\ Chinese, Turkish, Korean, Irish, Galician, Luxembourgish, Santali). New and updated translations are contributed on translatewiki.net and merged automatically, redeploying without developer intervention. This reduces linguistic barriers for international contributors and broadens commitment to linguistic diversity and equity. Moreover, by adopting SWHIDs as first-class identifiers for every indexed entity, the system renders each retrieved snippet permanently citable in academic work and documentation, operationalising the UNESCO Recommendation on Open Science \cite{UNESCO2022OpenScience}. Crucially, it does so at the granularity of individual software entities (each function, type, and template becoming a uniquely identified, referenceable artefact), advancing the broader agenda of reusable, well-referenced research software \cite{code-beyond-fair,dicosmo2025roadmap}. These persistent identifiers further allow archived entities to be reconciled into knowledge graphs, connecting code archiving to Wikidata through emerging software-citation and code-grounding workflows \cite{from_code_to_wikidata,cikm_pennino_2026}.

The service has been deployed on Wikimedia Toolforge (since April 2026). Its RESTful API, documented via Swagger UI at the \texttt{/docs} endpoint, facilitates programmatic integration into IDE plugins, bots, automated code-review workflows, and retrieval-augmented code generation \cite{coderag,coderag_bench}.

\section{Conclusions}

We have presented MediaWiki Code2Code Search, a neural retrieval system for the semantic discovery of structural software entities across the MediaWiki ecosystem. By leveraging Tree-sitter to index about 1.29 million functions, types, and templates, the system facilitates efficient navigation through 2,500+ repositories. Our implementation encodes these entities as 1,024-dimensional vectors using the \texttt{Qwen3-Embedding-0.6B} model. The resulting FAISS IVF-PQ index, occupying only 168.6 MiB, represents a 96.6\% reduction in memory compared to a flat \texttt{float32} baseline, enabling the system to serve nearest-neighbour queries on commodity CPU hardware with a median latency of 1.85 seconds.

The architecture employs a split-build paradigm that decouples GPU-intensive vectorisation from resource-constrained deployment. This enables the use of state-of-the-art neural models within the Wikimedia Toolforge environment while maintaining strict adherence to open-source infrastructure requirements.

Beyond its technical implementation, the system reinforces three core values: (i) open reproducibility, achieved by archiving all repositories in Software Heritage and attaching ISO/IEC 18670:2025-compliant SWHIDs to every result; (ii) linguistic equity, provided through a community-maintained localised interface (30 languages at the time of writing, 14 of which are Indian) continuously expanded via translatewiki.net; and (iii) ecosystem discoverability, through integration with Toolhub, Wikidata, and MediaWiki.org. While the current deployment focuses on MediaWiki, the underlying architecture is inherently modular and portable. Future research will explore cross-encoder reranking to enhance precision in critical workflows (extending the single-stage recall layer evaluated here into the recall-then-rerank pipeline \cite{two-stage-paradigm-2023}), implement incremental reindexing for near-real-time updates, and systematically evaluate the model's performance in supporting natural-language queries.

\section*{Funding}
This study was funded by the Alfred P. Sloan Foundation with the grant \#\href{https://sloan.org/grant-detail/g-2025-25193}{G-2025-25193} (\url{sloan.org}). It was also funded by Wikimedia CH via their micro-grant programme (\url{https://w.wiki/SrcE}).

\section*{Declaration of competing interests}
The authors declare the following financial interests/personal relationships which may be considered as potential competing interests:

Francesco Tosoni reports a relationship with Software Heritage (Inria Paris) that includes travel reimbursement. Francesco Tosoni reports a relationship with Wikimedia Foundation that includes travel reimbursement. In July 2025, the author conducted a research visit to Inria, supported by Software Heritage (SWH) through travel reimbursement, to discuss compression and code-search algorithms. Furthermore, the author is currently engaged in ongoing research collaborations with SWH regarding the development of archival and retrieval technologies. Additionally, the author has received a travel scholarship from the Wikimedia Foundation to present research related to software preservation at Wikimania 2026. These activities are academic and collaborative in nature and are not perceived to have influenced the integrity of the reported work. If there are other authors, they declare that they have no known competing financial interests or personal relationships that could have appeared to influence the work reported in this paper.

\section*{Acknowledgements}

All the offline code-vectorisation computations presented in this paper were performed using the GRICAD infrastructure (\href{https://gricad.univ-grenoble-alpes.fr}{https://gricad.univ-grenoble-alpes.fr}), which is supported by Grenoble research communities.

The author thanks Pierre Girard and the whole SOS Gricad team for technical support. He also thanks the Software Heritage (SWH) team in Paris and the members of CodeCommons WP3 for continuous technical feedback and discussions on this work; in particular, Erika Lena, Leonardo Venuta and Paolo Ferragina (Acube Lab, Pisa), and Gaël de Chalendar and Asma Graiess (CEA-List). 

Special thanks are due to the Wikimedia Toolforge community and administrators for providing the infrastructure, and to the members of the Indic MediaWiki Developers User Group and the Italian Wikimedia technical community for their insightful discussions and feedback.


\bibliographystyle{elsarticle-num} 
\bibliography{interactapasample}

\begin{thebibliography}{10}
\expandafter\ifx\csname url\endcsname\relax
  \def\url#1{\texttt{#1}}\fi
\expandafter\ifx\csname urlprefix\endcsname\relax\def\urlprefix{URL }\fi
\expandafter\ifx\csname href\endcsname\relax
  \def\href#1#2{#2} \def\path#1{#1}\fi

\bibitem{rasberry_github_wikidata_2023}
H.~Turki, M.~A.~H. Taieb, M.~B. Aouicha, L.~Rasberry, D.~Mietchen,
  \href{https://ceur-ws.org/Vol-3640/paper10.pdf}{Preregistration: Comparing
  the use of {Wikidata} and {Wikipedia} by open-source software programmers on
  {GitHub} repositories}, in: L.~Kaffee, S.~Razniewski, K.~Alghamdi, H.~Arnaout
  (Eds.), Proceedings of the Wikidata Workshop 2023 co-located with 22nd
  International Semantic Web Conference {(ISWC} 2023), Athens, Greece, November
  13, 2023, Vol. 3640 of {CEUR} Workshop Proceedings, CEUR-WS.org, 2023.
\newline\urlprefix\url{https://ceur-ws.org/Vol-3640/paper10.pdf}

\bibitem{working_mw_book}
Y.~Koren, \href{https://workingwithmediawiki.com/book/}{Working with
  MediaWiki}, 2nd Edition, WikiWorks Press, 2022, second edition, third
  printing.
\newline\urlprefix\url{https://workingwithmediawiki.com/book/}

\bibitem{toolkit-mining-wp}
D.~Milne, I.~H. Witten,
  \href{https://www.sciencedirect.com/science/article/pii/S000437021200077X}{An
  open-source toolkit for mining {Wikipedia}}, Artificial Intelligence 194
  (2013) 222--239, artificial Intelligence, Wikipedia and Semi-Structured
  Resources.
\newblock \href {https://doi.org/https://doi.org/10.1016/j.artint.2012.06.007}
  {\path{doi:https://doi.org/10.1016/j.artint.2012.06.007}}.
\newline\urlprefix\url{https://www.sciencedirect.com/science/article/pii/S000437021200077X}

\bibitem{wikidata-cacm}
D.~Vrande\v{c}i\'{c}, M.~Kr\"{o}tzsch, Wikidata: a free collaborative
  knowledgebase, Commun. ACM 57~(10) (2014) 78–85.
\newblock \href {https://doi.org/10.1145/2629489} {\path{doi:10.1145/2629489}}.

\bibitem{orcid_wd}
E.~Seidlmayer, J.~Vo{\ss}, T.~Melnychuk, L.~Galke, K.~Tochtermann, C.~Schultz,
  K.~U. F{\"{o}}rstner,
  \href{https://ceur-ws.org/Vol-2773/paper-09.pdf}{{ORCID} for {Wikidata}. data
  enrichment for scientometric applications}, in: L.~Kaffee,
  O.~Tifrea{-}Marciuska, E.~Simperl, D.~Vrandecic (Eds.), Proceedings of the
  1st Wikidata Workshop (Wikidata 2020) co-located with 19th International
  Semantic Web Conference(OPub 2020), Virtual Conference, November 2-6, 2020,
  {CEUR} Workshop Proceedings, CEUR-WS.org, 2020.
\newline\urlprefix\url{https://ceur-ws.org/Vol-2773/paper-09.pdf}

\bibitem{scholia_wd}
F.~{\AA}. Nielsen, D.~Mietchen, E.~Willighagen, {Scholia}, scientometrics and
  {Wikidata}, in: E.~Blomqvist, K.~Hose, H.~Paulheim, A.~{\L}awrynowicz,
  F.~Ciravegna, O.~Hartig (Eds.), The Semantic Web: ESWC 2017 Satellite Events,
  Springer International Publishing, Cham, 2017, pp. 237--259.
\newblock \href {https://doi.org/10.1007/978-3-319-70407-4_36}
  {\path{doi:10.1007/978-3-319-70407-4_36}}.

\bibitem{wd_wb_complementary}
L.~Rossenova, P.~Duchesne, I.~Bl{\"u}mel,
  \href{https://ceur-ws.org/Vol-3262/paper15.pdf}{{Wikidata} and {Wikibase} as
  complementary research data management services for cultural heritage data},
  in: Wikidata 2022 : Wikidata Workshop 2022, Proceedings of the 3rd Wikidata
  Workshop 2022 co-located with the 21st International Semantic Web Conference
  (ISWC2022), 2022, p.~15.
\newblock \href {https://doi.org/10.25968/opus-2573}
  {\path{doi:10.25968/opus-2573}}.
\newline\urlprefix\url{https://ceur-ws.org/Vol-3262/paper15.pdf}

\bibitem{code-search-co-attentive}
J.~Shuai, L.~Xu, C.~Liu, M.~Yan, X.~Xia, Y.~Lei, Improving code search with
  co-attentive representation learning, in: Proceedings of the 28th
  International Conference on Program Comprehension, ICPC '20, Association for
  Computing Machinery, New York, NY, USA, 2020, p. 196–207.
\newblock \href {https://doi.org/10.1145/3387904.3389269}
  {\path{doi:10.1145/3387904.3389269}}.

\bibitem{aroma_2019}
S.~Luan, D.~Yang, C.~Barnaby, K.~Sen, S.~Chandra, Aroma: code recommendation
  via structural code search, Proc. ACM Program. Lang. 3~(OOPSLA) (Oct. 2019).
\newblock \href {https://doi.org/10.1145/3360578} {\path{doi:10.1145/3360578}}.

\bibitem{mediawiki:codesearch}
\href{https://www.mediawiki.org/wiki/Codesearch}{{MediaWiki Codesearch}},
  MediaWiki, online; accessed 2026 (2026).
\newline\urlprefix\url{https://www.mediawiki.org/wiki/Codesearch}

\bibitem{hound}
{hound-search}, K.~Norton, J.~Klein,
  \href{https://github.com/hound-search/hound}{{Hound}: Lightning fast code
  searching made easy}, GitHub repository (2015).
\newline\urlprefix\url{https://github.com/hound-search/hound}

\bibitem{survey_llm_bigcode}
M.~Allamanis, E.~T. Barr, P.~Devanbu, C.~Sutton, A survey of machine learning
  for big code and naturalness, ACM Comput. Surv. 51~(4) (Jul. 2018).
\newblock \href {https://doi.org/10.1145/3212695} {\path{doi:10.1145/3212695}}.

\bibitem{aipowered-book}
T.~Grainger, D.~Turnbull, M.~Irwin, AI-Powered Search, Manning Publications,
  Shelter Island, NY, 2024, foreword by Grant Ingersoll.

\bibitem{deep-code-search}
X.~Gu, H.~Zhang, S.~Kim, Deep code search, in: 2018 IEEE/ACM 40th International
  Conference on Software Engineering (ICSE), 2018, pp. 933--944.
\newblock \href {https://doi.org/10.1145/3180155.3180167}
  {\path{doi:10.1145/3180155.3180167}}.

\bibitem{neural-code-search}
S.~Sachdev, H.~Li, S.~Luan, S.~Kim, K.~Sen, S.~Chandra, Retrieval on source
  code: a neural code search, in: Proceedings of the 2nd ACM SIGPLAN
  International Workshop on Machine Learning and Programming Languages, MAPL
  2018, Association for Computing Machinery, New York, NY, USA, 2018, p.
  31–41.
\newblock \href {https://doi.org/10.1145/3211346.3211353}
  {\path{doi:10.1145/3211346.3211353}}.

\bibitem{two-stage-paradigm-2023}
F.~Hu, Y.~Wang, L.~Du, X.~Li, H.~Zhang, S.~Han, D.~Zhang, Revisiting code
  search in a two-stage paradigm, in: Proceedings of the Sixteenth ACM
  International Conference on Web Search and Data Mining, WSDM '23, Association
  for Computing Machinery, New York, NY, USA, 2023, p. 994–1002.
\newblock \href {https://doi.org/10.1145/3539597.3570383}
  {\path{doi:10.1145/3539597.3570383}}.

\bibitem{code_search_large_scale}
J.~Martinez-Gil, S.~Yin, Evaluation of Code Similarity Search Strategies in
  Large-Scale Codebases, Springer Berlin Heidelberg, Berlin, Heidelberg, 2025,
  pp. 99--113.
\newblock \href {https://doi.org/10.1007/978-3-662-70140-9_4}
  {\path{doi:10.1007/978-3-662-70140-9_4}}.

\bibitem{wikidatasearch}
J.~Fraine, L.~Pintscher, P.~Saadé,
  \href{https://github.com/wmde/WikidataSearch}{{WikidataSearch: Retrieval
  Augmented Generation with the Wikidata Vector Database}}, web application:
  \url{https://wd-vectordb.wmcloud.org} (2026).
\newline\urlprefix\url{https://github.com/wmde/WikidataSearch}

\bibitem{wmf_ir_2026}
{Wikimedia Foundation Information Retrieval Working Group},
  \href{https://www.mediawiki.org/wiki/Readers/Information_Retrieval}{{Readers/Information
  Retrieval}}, MediaWiki project page (2026).
\newline\urlprefix\url{https://www.mediawiki.org/wiki/Readers/Information_Retrieval}

\bibitem{wise2023}
P.~Sridhar, H.~Lee, A.~Dutta, A.~Zisserman,
  \href{https://wikiworkshop.org/2023/papers/WikiWorkshop2023_paper_43.pdf}{{WISE}
  image search engine}, Wiki Workshop 2023, 10th edition, May 11, 2023; web
  application: \url{https://wise.wmcloud.org/} (May 2023).
\newline\urlprefix\url{https://wikiworkshop.org/2023/papers/WikiWorkshop2023_paper_43.pdf}

\bibitem{wise2026}
P.~Sridhar, H.~Lee, D.~M.~S. Pinto, A.~Zisserman, A.~Dutta, {WISE}: A
  multimodal search engine for visual scenes, audio, objects, faces, speech,
  and metadata, in: Proceedings of the 49th International ACM SIGIR Conference
  on Research and Development in Information Retrieval, SIGIR '26, Association
  for Computing Machinery, New York, NY, USA, 2026, p. 5230–5234.
\newblock \href {https://doi.org/10.1145/3805712.3808375}
  {\path{doi:10.1145/3805712.3808375}}.

\bibitem{swh-ecosystems-book}
T.~Mens, C.~De~Roover, A.~Cleve (Eds.), Software Ecosystems: Tooling and
  Analytics, 1st Edition, Springer Cham, 2023, xXII, 314 pages.
\newblock \href {https://doi.org/10.1007/978-3-031-36060-2}
  {\path{doi:10.1007/978-3-031-36060-2}}.

\bibitem{swh_cacm}
J.~Abramatic, R.~Di~Cosmo, S.~Zacchiroli, Building the universal archive of
  source code, Commun. {ACM} 61~(10) (2018) 29--31.
\newblock \href {https://doi.org/10.1145/3183558} {\path{doi:10.1145/3183558}}.

\bibitem{di-cosmo-archiving}
R.~Di~Cosmo, Archiving and referencing source code with {Software Heritage},
  in: A.~M. Bigatti, J.~Carette, J.~H. Davenport, M.~Joswig, T.~de~Wolff
  (Eds.), Mathematical Software -- ICMS 2020, Springer International
  Publishing, Cham, 2020, pp. 362--373.
\newblock \href {https://doi.org/10.1007/978-3-030-52200-1_36}
  {\path{doi:10.1007/978-3-030-52200-1_36}}.

\bibitem{zacchiroli-archiving-repro}
L.~Court\`{e}s, T.~Sample, S.~Zacchiroli, S.~Tournier, Source code archiving to
  the rescue of reproducible deployment, in: Proceedings of the 2nd ACM
  Conference on Reproducibility and Replicability, ACM REP '24, Association for
  Computing Machinery, New York, NY, USA, 2024, p. 36–45.
\newblock \href {https://doi.org/10.1145/3641525.3663622}
  {\path{doi:10.1145/3641525.3663622}}.

\bibitem{ISO-IEC-18670:2025}
\href{https://www.iso.org/standard/89985.html}{{ISO/IEC 18670:2025}.
  {I}nformation technology — {SoftWare Hash IDentifier} ({SWHID})
  specification {V1.2}} (April 2025).
\newline\urlprefix\url{https://www.iso.org/standard/89985.html}

\bibitem{swhid-website}
{SWHID Working Group}, \href{https://www.swhid.org}{{SWHID.org} — {SoftWare
  Hash IDentifier} ({SWHID})}, official website of the SWHID international
  standard (ISO/IEC 18670:2025) (2025).
\newline\urlprefix\url{https://www.swhid.org}

\bibitem{comparison_source_code}
C.~Niu, C.~Li, V.~Ng, D.~Chen, J.~Ge, B.~Luo, An empirical comparison of
  pre-trained models of source code, in: Proceedings of the 45th International
  Conference on Software Engineering, ICSE '23, IEEE Press, 2023, p.
  2136–2148.
\newblock \href {https://doi.org/10.1109/ICSE48619.2023.00180}
  {\path{doi:10.1109/ICSE48619.2023.00180}}.

\bibitem{qwen3_emb}
Y.~Zhang, M.~Li, D.~Long, X.~Zhang, H.~Lin, B.~Yang, P.~Xie, A.~Yang, D.~Liu,
  J.~Lin, F.~Huang, J.~Zhou, Qwen3 embedding: Advancing text embedding and
  reranking through foundation models (2025).
\newblock \href {http://arxiv.org/abs/2506.05176} {\path{arXiv:2506.05176}}.

\bibitem{bench_dl_code}
L.~Venuta, F.~Tosoni, P.~Ferragina, {Candidate Recall in Large-Scale
  Code-to-Code Retrieval: A Comprehensive Benchmark}, in: Web Information
  Systems Engineering -- WISE 2026, Springer, 2026, to appear.

\bibitem{sitter-wagner1998}
T.~A. Wagner, S.~L. Graham, Efficient and flexible incremental parsing, ACM
  Trans. Program. Lang. Syst. 20~(5) (1998) 980–1013.
\newblock \href {https://doi.org/10.1145/293677.293678}
  {\path{doi:10.1145/293677.293678}}.

\bibitem{sitter-comparison2023}
A.~Latif, F.~Azam, M.~W. Anwar, A.~Zafar, Comparison of leading language
  parsers – {ANTLR}, {JavaCC}, {SableCC}, {Tree-sitter}, {Yacc}, {Bison}, in:
  2023 13th International Conference on Software Technology and Engineering
  (ICSTE), 2023, pp. 7--13.
\newblock \href {https://doi.org/10.1109/ICSTE61649.2023.00009}
  {\path{doi:10.1109/ICSTE61649.2023.00009}}.

\bibitem{sitter-agnostic2025}
F.~Refolli, D.~Sas, F.~A. Fontana, Lessons learned from implementing a
  language-agnostic dependency graph parser, in: Proceedings of the 20th
  International Conference on Evaluation of Novel Approaches to Software
  Engineering - ENASE, INSTICC, SciTePress, 2025, pp. 484--491.
\newblock \href {https://doi.org/10.5220/0013277600003928}
  {\path{doi:10.5220/0013277600003928}}.

\bibitem{sentence2019}
N.~Reimers, I.~Gurevych,
  \href{https://aclanthology.org/D19-1410/}{Sentence-{BERT}: Sentence
  embeddings using {S}iamese {BERT}-networks}, in: K.~Inui, J.~Jiang, V.~Ng,
  X.~Wan (Eds.), Proceedings of the 2019 Conference on Empirical Methods in
  Natural Language Processing and the 9th International Joint Conference on
  Natural Language Processing (EMNLP-IJCNLP), Association for Computational
  Linguistics, Hong Kong, China, 2019, pp. 3982--3992.
\newblock \href {https://doi.org/10.18653/v1/D19-1410}
  {\path{doi:10.18653/v1/D19-1410}}.
\newline\urlprefix\url{https://aclanthology.org/D19-1410/}

\bibitem{FAISS}
M.~Douze, A.~Guzhva, C.~Deng, J.~Johnson, G.~Szilvasy, P.-E. Mazaré,
  M.~Lomeli, L.~Hosseini, H.~Jégou, The {Faiss} library, IEEE Transactions on
  Big Data 12~(2) (2026) 346--361.
\newblock \href {https://doi.org/10.1109/TBDATA.2025.3618474}
  {\path{doi:10.1109/TBDATA.2025.3618474}}.

\bibitem{prodquant}
H.~Jégou, M.~Douze, C.~Schmid, Product quantization for nearest neighbor
  search, IEEE Transactions on Pattern Analysis and Machine Intelligence 33~(1)
  (2011) 117--128.
\newblock \href {https://doi.org/10.1109/TPAMI.2010.57}
  {\path{doi:10.1109/TPAMI.2010.57}}.

\bibitem{code_search_python_translation}
M.~M. Khan, Z.~Yu, Approaching code search for python as a translation
  retrieval problem with dual encoders, Empirical Software Engineering 30~(1)
  (2024) 12.
\newblock \href {https://doi.org/10.1007/s10664-024-10580-3}
  {\path{doi:10.1007/s10664-024-10580-3}}.

\bibitem{hnsw}
Y.~A. Malkov, D.~A. Yashunin, Efficient and robust approximate nearest neighbor
  search using hierarchical navigable small world graphs, IEEE Transactions on
  Pattern Analysis and Machine Intelligence 42~(4) (2020) 824--836.
\newblock \href {https://doi.org/10.1109/TPAMI.2018.2889473}
  {\path{doi:10.1109/TPAMI.2018.2889473}}.

\bibitem{pan2024vector-db}
J.~J. Pan, J.~Wang, G.~Li, Vector database management techniques and systems,
  in: Companion of the 2024 Int. Conf. on Management of Data, SIGMOD '24,
  Assoc. for Computing Machinery, New York, NY, USA, 2024, p. 597–604.
\newblock \href {https://doi.org/10.1145/3626246.3654691}
  {\path{doi:10.1145/3626246.3654691}}.

\bibitem{pitfalls_vector_sigmod26}
T.~Chen, C.~Fu, J.~Wu, H.~Wu, H.~Fan, X.~Ke, Y.~Gao, Y.~Ni, A.~Zeng, Reveal
  hidden pitfalls and navigate next generation of vector similarity search from
  task-centric views: [experiments \& analysis], Proc. ACM Manag. Data 4~(1)
  (Apr. 2026).
\newblock \href {https://doi.org/10.1145/3786690} {\path{doi:10.1145/3786690}}.

\bibitem{gaurav-survey2023}
G.~Menghani, Efficient deep learning: A survey on making deep learning models
  smaller, faster, and better, ACM Comput. Surv. 55~(12) (Mar. 2023).
\newblock \href {https://doi.org/10.1145/3578938} {\path{doi:10.1145/3578938}}.

\bibitem{codemeta}
M.~B. Jones, C.~Boettiger, A.~{Cabunoc Mayes}, A.~Smith, M.~Gruenpeter,
  V.~Lorentz, T.~Morrell, D.~Garijo, P.~Slaughter, K.~Niemeyer, Y.~Gil,
  M.~Fenner, K.~Nowak, M.~Hahnel, L.~Coy, A.~Allen, M.~Crosas, A.~Sands,
  N.~{Chue Hong}, P.~Cruse, D.~S. Katz, C.~Goble, B.~Mecum,
  A.~Gonzalez-Beltran, N.~Ross,
  \href{https://w3id.org/codemeta/3.0}{{CodeMeta}: an exchange schema for
  software metadata}, version 3.0 (2023).
\newline\urlprefix\url{https://w3id.org/codemeta/3.0}

\bibitem{codesearchnet}
H.~Husain, H.-H. Wu, T.~Gazit, M.~Allamanis, M.~Brockschmidt, {CodeSearchNet}
  challenge: Evaluating the state of semantic code search (2020).
\newblock \href {http://arxiv.org/abs/1909.09436} {\path{arXiv:1909.09436}}.

\bibitem{xCodeEval}
M.~A.~M. Khan, M.~S. Bari, X.~L. Do, W.~Wang, M.~R. Parvez, S.~Joty,
  \href{https://aclanthology.org/2024.acl-long.367/}{{XC}ode{E}val: An
  execution-based large scale multilingual multitask benchmark for code
  understanding, generation, translation and retrieval}, in: L.-W. Ku,
  A.~Martins, V.~Srikumar (Eds.), Proceedings of the 62nd Annual Meeting of the
  Association for Computational Linguistics (Volume 1: Long Papers),
  Association for Computational Linguistics, Bangkok, Thailand, 2024, pp.
  6766--6805.
\newblock \href {https://doi.org/10.18653/v1/2024.acl-long.367}
  {\path{doi:10.18653/v1/2024.acl-long.367}}.
\newline\urlprefix\url{https://aclanthology.org/2024.acl-long.367/}

\bibitem{voorhees1998trecpool}
E.~M. Voorhees, Variations in relevance judgments and the measurement of
  retrieval effectiveness, in: Proceedings of the 21st Annual International ACM
  SIGIR Conference on Research and Development in Information Retrieval, SIGIR
  '98, Association for Computing Machinery, New York, NY, USA, 1998, p.
  315–323.
\newblock \href {https://doi.org/10.1145/290941.291017}
  {\path{doi:10.1145/290941.291017}}.

\bibitem{ir_book}
C.~D. Manning, P.~Raghavan, H.~Schütze, Introduction to Information Retrieval,
  Cambridge University Press, 2008.
\newblock \href {https://doi.org/10.1017/CBO9780511809071}
  {\path{doi:10.1017/CBO9780511809071}}.

\bibitem{robertson1994okapi}
S.~E. Robertson, S.~Walker, S.~Jones, M.~Hancock{-}Beaulieu, M.~Gatford,
  \href{http://trec.nist.gov/pubs/trec3/papers/city.ps.gz}{Okapi at {TREC-3}},
  in: D.~K. Harman (Ed.), Proceedings of The Third Text REtrieval Conference,
  {TREC} 1994, Gaithersburg, Maryland, USA, November 2-4, 1994, {NIST} Special
  Publication, National Institute of Standards and Technology {(NIST)}, 1994,
  pp. 109--126.
\newline\urlprefix\url{http://trec.nist.gov/pubs/trec3/papers/city.ps.gz}

\bibitem{zheng2023judging}
L.~Zheng, W.-L. Chiang, Y.~Sheng, S.~Zhuang, Z.~Wu, Y.~Zhuang, Z.~Lin, Z.~Li,
  D.~Li, E.~P. Xing, H.~Zhang, J.~E. Gonzalez, I.~Stoica, Judging
  {LLM}-as-a-judge with {MT}-bench and {Chatbot Arena}, in: Proceedings of the
  37th International Conference on Neural Information Processing Systems, NIPS
  '23, Curran Associates Inc., Red Hook, NY, USA, 2023.

\bibitem{sacl-textual-bias}
D.~Gupta, G.~G. Lakshmy, Y.~Xie,
  \href{https://aclanthology.org/2025.findings-emnlp.1365/}{{SACL}:
  Understanding and combating textual bias in code retrieval with
  semantic-augmented reranking and localization}, in: C.~Christodoulopoulos,
  T.~Chakraborty, C.~Rose, V.~Peng (Eds.), Findings of the Assoc. for
  Computational Linguistics, Assoc. for Computational Linguistics, Suzhou,
  China, 2025, pp. 25052--25065.
\newblock \href {https://doi.org/10.18653/v1/2025.findings-emnlp.1365}
  {\path{doi:10.18653/v1/2025.findings-emnlp.1365}}.
\newline\urlprefix\url{https://aclanthology.org/2025.findings-emnlp.1365/}

\bibitem{sourcerercc}
H.~Sajnani, V.~Saini, J.~Svajlenko, C.~K. Roy, C.~V. Lopes, {SourcererCC}:
  scaling code clone detection to big-code, in: Proceedings of the 38th
  International Conference on Software Engineering, ICSE '16, Association for
  Computing Machinery, New York, NY, USA, 2016, p. 1157–1168.
\newblock \href {https://doi.org/10.1145/2884781.2884877}
  {\path{doi:10.1145/2884781.2884877}}.

\bibitem{bigclone}
J.~Svajlenko, J.~F. Islam, I.~Keivanloo, C.~K. Roy, M.~M. Mia, Towards a big
  data curated benchmark of inter-project code clones, in: Proceedings of the
  2014 IEEE International Conference on Software Maintenance and Evolution,
  ICSME '14, IEEE Computer Society, USA, 2014, p. 476–480.
\newblock \href {https://doi.org/10.1109/ICSME.2014.77}
  {\path{doi:10.1109/ICSME.2014.77}}.

\bibitem{swh_dag}
A.~Pietri, D.~Spinellis, S.~Zacchiroli, The {Software Heritage} graph dataset:
  Large-scale analysis of public software development history, in: Proceedings
  of the 17th International Conference on Mining Software Repositories, MSR
  '20, Association for Computing Machinery, New York, NY, USA, 2020, p. 1–5.
\newblock \href {https://doi.org/10.1145/3379597.3387510}
  {\path{doi:10.1145/3379597.3387510}}.

\bibitem{UNESCO2022OpenScience}
{UNESCO}, {Canadian Commission for UNESCO},
  \href{https://unesdoc.unesco.org/ark:/48223/pf0000383771}{An introduction to
  the {UNESCO} recommendation on open science}, Programme and meeting document
  SC-PBS-STIP/2022/OSB/1, UNESCO, licensed under CC BY-SA 3.0 IGO. (2022).
\newblock \href {https://doi.org/10.54677/XOIR1696}
  {\path{doi:10.54677/XOIR1696}}.
\newline\urlprefix\url{https://unesdoc.unesco.org/ark:/48223/pf0000383771}

\bibitem{code-beyond-fair}
R.~Di~Cosmo, S.~Granger, K.~Hinsen, N.~Jullien, D.~Le~Berre, V.~Louvet,
  C.~Maumet, C.~Maurice, R.~Monat, N.~P. Rougier, Code beyond {FAIR}: a roadmap
  for reusable research software, Scientific Data 13~(1) (2026) 514.
\newblock \href {https://doi.org/10.1038/s41597-026-06705-6}
  {\path{doi:10.1038/s41597-026-06705-6}}.

\bibitem{dicosmo2025roadmap}
R.~Di~Cosmo, S.~Granger, K.~Hinsen, N.~Jullien, D.~Le~Berre, V.~Louvet,
  C.~Maumet, C.~Maurice, R.~Monat, N.~P. Rougier, Stop treating code like an
  afterthought: record, share and value it, Nature 646~(8084) (2025) 284--286.
\newblock \href {https://doi.org/10.1038/d41586-025-03196-0}
  {\path{doi:10.1038/d41586-025-03196-0}}.

\bibitem{from_code_to_wikidata}
C.~C. Pellizzari~di San~Girolamo, F.~Tosoni, From code archival to knowledge
  graph: Bridging {Software Heritage}, {COAR Notify} and {Wikidata}, in:
  Proceedings of the Wikidata Workshop 2026 co-located with ISWC 2026, 2026, to
  appear.

\bibitem{cikm_pennino_2026}
F.~Pennino, A.~Gurioli, S.~Zacchiroli, M.~Gabbrielli, P.~Ferragina, Codegraph:
  Open-taxonomy knowledge graph for source code with wikidata grounding, in:
  Proceedings of the 35th ACM International Conference on Information and
  Knowledge Management (CIKM '26), ACM, Rome, Italy, 2026, to appear.
\newblock \href {https://doi.org/10.1145/3799682.3840769}
  {\path{doi:10.1145/3799682.3840769}}.

\bibitem{coderag}
S.~Zhang, Y.~Ding, S.~Lian, S.~Song, H.~Li,
  \href{https://aclanthology.org/2025.emnlp-main.1187/}{{C}ode{RAG}: Finding
  relevant and necessary knowledge for retrieval-augmented repository-level
  code completion}, in: C.~Christodoulopoulos, T.~Chakraborty, C.~Rose, V.~Peng
  (Eds.), Proceedings of the 2025 Conference on Empirical Methods in Natural
  Language Processing, Association for Computational Linguistics, Suzhou,
  China, 2025, pp. 23278--23288.
\newblock \href {https://doi.org/10.18653/v1/2025.emnlp-main.1187}
  {\path{doi:10.18653/v1/2025.emnlp-main.1187}}.
\newline\urlprefix\url{https://aclanthology.org/2025.emnlp-main.1187/}

\bibitem{coderag_bench}
Z.~Z. Wang, A.~Asai, X.~V. Yu, F.~F. Xu, Y.~Xie, G.~Neubig, D.~Fried,
  \href{https://aclanthology.org/2025.findings-naacl.176/}{{C}ode{RAG}-bench:
  Can retrieval augment code generation?}, in: L.~Chiruzzo, A.~Ritter, L.~Wang
  (Eds.), Findings of the Association for Computational Linguistics: NAACL
  2025, Association for Computational Linguistics, Albuquerque, New Mexico,
  2025, pp. 3199--3214.
\newblock \href {https://doi.org/10.18653/v1/2025.findings-naacl.176}
  {\path{doi:10.18653/v1/2025.findings-naacl.176}}.
\newline\urlprefix\url{https://aclanthology.org/2025.findings-naacl.176/}

\end{thebibliography}

\end{document}